\documentclass[notoc]{JHEP}

\usepackage{graphicx}
\usepackage{dcolumn}
\usepackage{bm}
\input mathrsfs.sty

\def\Tr{\mbox{Tr}\,}

\def\D{{\mathscr D}}
\def\half{\frac{1}{2}}
\def\d{\partial}
\def\m{\mu}
\def\n{\nu}
\def\slash#1{\, /\kern-0.6em{#1}}

\title{Monopoles and flux strings from SU(2) adjoint scalars}

\author{Chandrasekhar Chatterjee, Amitabha Lahiri\\
\email{chandra@bose.res.in}, \email{amitabha@bose.res.in}\\ 
Department of Theoretical Sciences\\ 
S.~N.~Bose National Centre for Basic Sciences\\ 
Block JD, Sector III, Salt Lake, Kolkata 700 098, W.B. India. }

\abstract{We construct, in an SU(2) gauge theory with two adjoint scalars,
flux strings with monopoles attached at the ends. One scalar breaks
SU(2) $\to$ U(1) and produces monopoles, the other then breaks the
U(1) and produces strings. Dualizing, we write the theory in terms
of effective string variables and show that the flux in the string
is exactly saturated by the monopoles at the ends.}

\begin{document}


\section{\label{intro}Introduction}
Strong interactions are adequately described at high energies by
quantum chromodynamics (QCD). At low energies, the QCD coupling is
large and color is confined, but a precise description of how that
happens is as yet unknown. Following the ideas of Mandelstam and
Nambu~\cite{Mandelstam:1974pi, Nambu:1975ba}, it is now generally
thought that the QCD vacuum behaves like a dual superconductor,
created by condensation of magnetic monopoles, in which confinement
is analogous to a dual Meissner effect. A meson state is then
formed by attaching quarks to the ends of a flux string analogous
to the Abrikosov-Nielsen-Olesen vortex string of Abelian gauge
theory~\cite{Abrikosov:1956sx, Nielsen:1973cs}.

A construction of flux strings in the Weinberg-Salam theory was
suggested by Nambu~\cite{Nambu:1977ag}, in which a pair of magnetic
monopoles are bound by a flux string of Z condensate. A different
construction of flux string, involving two adjoint scalar fields in
an SU(2) gauge theory, has been discussed in~\cite{Nielsen:1973cs,
  deVega:1976rt}. Recently there has been a resurgence of interest
in such constructions~\cite{Auzzi:2003fs, Hanany:2004ea,
  Shifman:2002yi, 'tHooft:1999au}.

In this paper we construct these strings and write the action in
terms of string variables as a dual gauge theory. A pair of 't
Hooft-Polyakov magnetic monopole and
anti-monopole~\cite{'tHooft:1974qc, Polyakov:1974ek, Prasad:1975kr}
are attached to the ends of the string, as in the picture of
confinement. We start with SU(2) gauge theory with two adjoint
scalars. One of them, call it $\phi_1\,,$ acquires a vacuum
expectation value (vev) $\vec v_1$ which is a vector in internal
space, and breaks the symmetry group down to U(1). The 't
Hooft-Polyakov monopoles are associated with this breaking. The
other scalar field $\phi_2$ also has a non-vanishing vev $\vec
v_2$, which is free to wind around $\vec v_1$ in the internal
space. This winding is mapped to a circle in space, giving rise to
the vortex string. We then dualize the fields as
in~\cite{Davis:1988rw, Mathur:1991ip, Lee:1993ty, Akhmedov:1995mw,
  Chatterjee:2006iq} to write the action in terms of string
variables.

The idea of two-scale symmetry breaking in SU(2), the first to
produce monopoles and the second to produce strings, has appeared
in~\cite{Hindmarsh:1985xc}. Later this idea was used in a
supersymmetric setting in~\cite{Kneipp:2003ue, Auzzi:2003em,
  Eto:2006dx}. The idea of flux matching, following
Nambu~\cite{Nambu:1977ag} also appeared in these papers. In this
paper we describe an explicit construction of flux strings in
non-supersymmetric SU(2) theory with 't Hooft-Polyakov monopoles of
the same theory attached to the ends. We also show explicitly, by
dualizing the variables, that the flux at each end of the string is
saturated by the magnetic monopoles, indicating confinement of
magnetic flux.  As far as we are aware, this is the first time this
construction has been done in terms of the variables corresponding
to the macroscopic objects -- the monopoles and flux tubes.

\section{Magnetic monopoles}
In the bulk of the paper we shall consider SU(2) gauge theory with
two adjoint scalar fields.  The Lagrangian for SU(2) gauge theory
with two adjoint scalar can be written as
\begin{eqnarray}
\label{lagrangian_1}
{\mathcal L} = - {\frac 12}\Tr\left(G_{\mu\nu}G^{\mu\nu}\right) + 
\Tr\left(D_\mu \phi_1 D^\mu \phi_1
\right) + \Tr\left(D_\mu \phi_2D^\mu \phi_2\right)  + V(\phi_1, \phi_2).
\end{eqnarray} 
Here $\phi_\alpha = \phi_\alpha^i \tau^i, (\alpha = 1,2)$ are two
real scalar fields in the adjoint representation of $SU(2)$,
$\tau^i$ are the $SU(2)$ generators with $\Tr(\tau^i\tau^j) =
\half\delta^{ij}\, (i,j = 1,2,3)$, and $V(\phi_1, \phi_2)$ is some
interaction potential for the scalars.  The covariant derivative
and the Yang-Mills field strength tensor are defined as
\begin{eqnarray}
{D_\mu{\phi_\alpha}}^i &=& \partial_\mu \phi_\alpha^i 
+ g \epsilon^{ijk}A_{\mu}^j\phi_\alpha^k ,\\
G_{\mu\nu}^i &=& \partial_\mu {A^i}_{\nu}- \partial_\nu {A^i}_{\mu}  
+ g \epsilon^{ijk}{A^j}_{\mu}{A^k}_{\nu}.
\end{eqnarray}
We will sometimes employ vector notation, in which
\begin{eqnarray}
D_\mu\vec\phi_\alpha &=& \partial_\mu\vec\phi_\alpha + g\vec A
\times \vec\phi_\alpha\,, \\
\vec G_{\mu\nu}&=& \partial_\mu\vec A_\nu -
\partial_\nu \vec A_\mu + g \vec A_\mu \times\vec A_\nu\,,\, {\rm
  etc.} 
\end{eqnarray}
Obviously, $\vec\phi_1$ and $\phi_1$ represent the same object.  The
simplest form of the potential $V(\phi_1, \phi_2)$ that will serve
our purpose is
\begin{eqnarray}
\label{potential}
V(\phi_1, \phi_2) = - {\frac{\lambda_1} 4}(|\phi_1|^2 -
v_1^2)^2 - {\frac{\lambda_2} 4}(|\phi_2|^2 - v_2^2)^2
-V_{mix}(\phi_1,\phi_2).  
\end{eqnarray}
The last term $V_{mix}(\phi_1,\phi_2)$ includes all mixing terms in
the potential, which involve products of the two scalar fields in
some way. Through an obvious abuse of terminology, we will refer to
$v_1,v_2$ as the vacuum expectation values of $\phi_1$ and
$\phi_2$, although in general the potential $V(\phi_1, \phi_2)$
will not reach local minima at $v_1,v_2.$ For now we will set
$V_{mix}(\phi_1,\phi_2) = 0$, then $v_1, v_2$ will in fact define
the local minima of the potential.  Let us now define the $\phi_1$
vacuum by
\begin{eqnarray}
\label{1sthiggsvacuum}
D_\mu{\vec\phi_1} = 0\\
|\phi_1|^2 = v_1^2.\nonumber
\end{eqnarray}
The vacuum expectation value of the adjoint field $\phi_1$ breaks
the SU(2) symmetry down to U(1) at the scale $v_1$.  The vector
fields $A^\pm$ and the modulus of the scalar field $\phi_1$ will
acquire a mass,
\begin{eqnarray}
M_{A^\pm} = gv_1, \qquad M_{|\phi_1|} = \sqrt \lambda_1 v_1.
\end{eqnarray}
We assume that the vacuum expectation value $v_1$ is large compared
to the energy scale we are interested in. This could happen, for
example, if the model is derived from a supersymmetric theory in
which the symmetry breaking occurs at a SUSY GUT scale. The large
value of $v_1$ provides large masses to the gauge fields and the
scalar. Well below the scale $v_1$ the massive gauge fields will
not appear and the scalar field will be at its vacuum value
$v_1$. So at low energies the theory is essentially Abelian, with
one massless gauge field, which we call $B_\mu$.  

At the $\phi_1$-vacuum the gauge field can be written
as~\cite{Corrigan:1975hd}, 
\begin{eqnarray}
\label{A_1sthiggsvacuum}
\vec A_\mu = B_{\mu}\hat{\phi}_1 - {\frac 1g}
\hat{\phi}_1\times\d_{\mu} \hat{\phi}_1\,,
\end{eqnarray}
where $B_{\mu} = \vec A_{\mu}\cdot \hat{\phi}_1$ and $\hat\phi_1 =
\vec\phi_1/v_1.$ Below the SU(2) breaking scale $v_1$, the gauge
field is defined by eq.~(\ref{A_1sthiggsvacuum}).  Looking at
eq.~(\ref{A_1sthiggsvacuum}) we can see that one of the three
components of the gauge field is not fully determined after
symmetry breaking.  It is a massless mode, which we will call
$B_\mu$.  The massive modes ${A_\mu}^\pm$ do not appear here
because they are not excited below the scale $v_1$. In a
configuration with the scalar field as $\phi_1^i \to
v_1\displaystyle{\frac {r^i}{r}}$ at spatial infinity, the second
term of the right hand side of eq.~(\ref{A_1sthiggsvacuum})
corresponds to the non-Abelian gauge field for magnetic
monopoles~\cite{'tHooft:1999au}.
 
Using Eq.~(\ref{A_1sthiggsvacuum}) we can write down the field
strength tensor at the $\phi_1$-vacuum,
\begin{eqnarray}
\vec{G}_{\mu\nu} = \left[\d_\mu B_\nu - \d_\nu B_\mu + {\frac
    1g}\hat{\phi}_1\cdot\d_\mu \hat{\phi}_1\times
  \d_\nu\hat{\phi}_1] \right] \hat\phi_1 - {\frac
  {2}{g}}\d_\mu\hat{\phi}_1\times\d_\nu \hat{\phi}_1.  
\end{eqnarray}
A straightforward calculation shows that,
\begin{eqnarray}
-{\frac 12}\Tr (G_{\mu\n}G^{\mu\n}) &=& -{\frac
  14}F_{\mu\n}F^{\mu\n},
\end{eqnarray}
where
\begin{eqnarray}
\label{F}
F_{\mu\n} &=& \d_{\mu}B_{\n} - \d_\nu B_\mu - 
{\frac 1g}\hat{\phi}_1\cdot\d_\mu \hat{\phi}_1\times
\d_\nu\hat{\phi}_1 
\end{eqnarray}
The second term of eq.~(\ref{F}) is the `monopole term'.  In a
configuration where the scalar field at spatial infinity goes as
$\phi_1^i \to v_1\displaystyle{\frac {r^i}{r}}$, the $(ij)^{th}$
component of the second term of eq.~(\ref{F}) becomes
$-\displaystyle{{\frac{\epsilon_{ijk} r^k}{gr^3}}},$ which we can
easily identify as the field of a magnetic monopole.  The flux for
this monopole field is ${4\pi}\over g $.  If the monopole has
magnetic charge $Q_m,$ its flux is $4\pi Q_m,$ and thus we find the
quantization condition for unit charge, $Q_m g = 1.$

The Lagrangian at the $\phi_1$-vacuum becomes
\begin{eqnarray}
{\mathcal L} = -  {\frac 14}F_{\mu\nu}F^{\mu\nu} + \half D_\mu
\vec\phi_2 \cdot D^\mu \vec\phi_2 
 -  {\frac{\lambda_2} 4}(|\phi_2|^2 - v_2^2)^2,
\label{vaclag}
\end{eqnarray}
with $F_{\mu\nu}$ given by eq.~(\ref{F}). 

\section{$\phi_2$ string in the $\phi_1$ vacuum}
We started with a theory with SU(2) symmetry and a pair of adjoint
scalars.  The non-zero vacuum expectation value of the field
$\phi_1$ breaks the symmetry to U(1) at a scale $v_1$.  Below $v_1$
the theory is effectively an Abelian theory with magnetic
monopoles.  The vacuum was chosen according to
eq.~(\ref{1sthiggsvacuum}).  However we did not fix the internal
direction of the field $\phi_1$.  This is because of the fact that
the gauge group $SU(2)$ is transitive on the vacuum manifold $S^2$
and this transitivity makes the theory irrespective of the
direction of $\phi_1$.  There is a little group U(1) in the theory
which leaves $\phi_1$ invariant on the vacuum and this little group
becomes the remaining symmetry of the theory. Actually the little
group is defined at every point on the vacuum. However, for our
case the little group action is the rotation around any point on
the vacuum manifold $S^2$. That is why, the little group is same
for every point on $S^2$ and the little group is U(1). So the
remaining symmetry is also independent of the direction of
$\phi_1$.

However, there is another scalar field in the theory, namely
$\phi_2$.  This $\phi_2$ is in the adjoint representation of SU(2),
so it has three real scalar components.  Out of the three, one
component can be chosen along the direction of the field $\phi_1.$
Then the other two will rotate on a two dimensional plane normal to
$\phi_1$ under the action of the little group U(1).  In other
words, a gauge transformation rotates $\phi_2$ around $\phi_1$.
This $U(1)$ will produce flux tubes when the $U(1)$ symmetry is
spontaneously broken down to $Z_2$.  It is natural to take the
$U(1)$ breaking scale $v_2$ to be very small compared to the
$SU(2)$ symmetry breaking scale, $v_2\ll v_1\,.$

In order to find string configurations, we write
the covariant derivative of $\phi_2$ using
eq.~(\ref{A_1sthiggsvacuum}),
\begin{eqnarray}
D_\mu\vec\phi_2 &=& \d_\mu\vec\phi_2 + g \vec
A_{\mu}\times\vec\phi_2,\nonumber \\ 
&=& \d_\mu \vec\phi_2 + g \left[B_{\mu}{\hat\phi_1 } - 
{\frac 1g}
\hat\phi_1\times{\d_{\mu}{\hat\phi_1}}\right]\times\vec\phi_2 
,\nonumber\\ 
\label{dmuphi}
&=&{\d_\mu \vec\phi_2} + g B_{\mu}{\hat\phi_1}\times\vec\phi_2 +
\left[\hat\phi_1\left(\d_\mu \hat\phi_1\cdot\vec\phi_2\right) -
  \d_\mu \hat\phi_1  \left(\hat\phi_1\cdot\vec\phi_2\right)
\right] .
\end{eqnarray}
This is of course in the $\phi_1$ vacuum.

For string configurations, $\phi_2$ has to approach its vacuum
value at far away from the string ($\phi_1$ is of course at the
$\phi_1$-vacuum already). The  $\phi_2$ vacuum is defined by 
\begin{eqnarray}
\label{2ndhiggs1}
|\vec{\phi_2}|^2 = v_2^2\,,\\
\label{2ndhiggs2}
D_\mu \vec{\phi_2} = 0.
\end{eqnarray}
These equations are taken in the $\phi_1$ vacuum, so in particular
we use Eq.~(\ref{dmuphi}) in the place of Eq.~(\ref{2ndhiggs2}). If
we now dot both sides of eq.~(\ref{2ndhiggs2}) with $ \hat\phi_1$,
we get
\begin{eqnarray}
\label{phi1phi2}
\d_\mu(\vec\phi_1\cdot\vec\phi_2) = 0.
\label{phidotphi}
\end{eqnarray}
So in the $\phi_2$ vacuum (which by definition is embedded in the
$\phi_1$ vacuum), the component of $\vec\phi_2$ along $\vec\phi_1$
remains constant.  

As mentioned above, we can decompose $\phi_2$ (not necessarily in
the $\phi_2$ vacuum) into a component along $\phi_1$ and another
component normal to $\phi_1$ in the internal space,
\begin{eqnarray}
\vec\phi_2 = (\hat\phi_1\cdot\vec\phi_2)\hat\phi_1 + \vec K\,. 
\label{defK}
\end{eqnarray}
Then 
\begin{eqnarray}
D_\mu\vec\phi_2 = \hat\phi_1\d_\mu(\hat\phi_1\cdot\vec\phi_2) +
\d_\mu\vec K - \hat\phi_1(\hat\phi_1\cdot\d_\mu\vec K) +
gB_\mu\hat\phi_1\times\vec K\,, 
\end{eqnarray}
and
\begin{eqnarray}
(D_\mu\vec\phi_2)^2 = \big(\d_\mu(\hat\phi_1\cdot\vec\phi_2)\big)^2
+ g^2(B_\mu \vec K)^2
+ (\d_\mu\vec K)^2 - (\hat\phi_1\cdot \d_\mu\vec
K)^2 + 2g B_\mu\hat\phi_1\times\vec K\cdot\d^\mu\vec K\,.\nonumber
\\
\end{eqnarray}
Since $\hat\phi_1\cdot\vec K = 0,$ we can rewrite this after a
little manipulation as 
\begin{eqnarray}
(D_\mu\vec\phi_2)^2 = \big(\d_\mu(\hat\phi_1\cdot\vec\phi_2)\big)^2  
+ (\partial_\mu k)^2 + k^2\left(\hat\kappa \cdot \d_\mu\hat \kappa
  \times \hat \phi_1 +   gB_\mu\right)^2\,,
\label{dphi2sq}
\end{eqnarray}
where we have written $\vec K = k\hat\kappa\,.$ We put this
expression into the action of Eq.~(\ref{vaclag}). Then in order to
extract the string variables, we note that at infinite distance
away from the string, $\phi_2$ approaches its vacuum value
$|\phi_2| \to v_2$. Further, according to Eq.~(\ref{phidotphi}),
$\hat\phi_1\cdot\vec\phi_2$ also approaches a constant, so using
Eq.~(\ref{defK}) we see that $k$ should also approach a constant. 
Then the first two terms of Eq.~(\ref{dphi2sq}) disappear at
infinity, as does the last term of Eq.~(\ref{vaclag}), and the
Lagrangian at infinity behaves as 
\begin{eqnarray}
{\mathcal L} =  {\frac {k^2} 2} \left(\hat \kappa \cdot
  \d_\mu\hat \kappa   \times \hat \phi_1 + gB_\mu\right)^2 - {\frac
  14} F^{\mu\n}F_{\mu\n},
\label{vaclag2}
\end{eqnarray}
where now $k$ is a constant. 

Since $|\vec\phi_2|$ and the component of $\vec\phi_2$ along
$\hat\phi_1$ both approach constant values at infinity, and so does
$k = |\vec K|,$ the only degree of freedom remaining in
$\vec\phi_2$ at infinity is an angle $\chi$ which parametrizes the
rotation of $\vec\phi_2$ around $\hat\phi_1.$ The first term inside
the brackets in Eq.~(\ref{vaclag2}) provides the $\partial_\mu\chi$
as we will see below.  This is the angle which is mapped onto a
circle at infinity to produce a flux string. Further, the system is
in the $\phi_1$-vacuum, i.e. $\vec\phi_1$ is in a vacuum
configuration given by Eqs.~(\ref{1sthiggsvacuum}) and
(\ref{A_1sthiggsvacuum}). So in particular we can choose this
vacuum to contain 't~Hooft-Polyakov monopoles as discussed after
Eq.~(\ref{A_1sthiggsvacuum}).

\section{Monopoles and Strings}
With the above in mind, let us parametrize the $\phi_1$-vacuum as
\begin{eqnarray}
\label{Uphi}
\vec\phi_1 = v_1 U(\vec x)\tau_3U(\vec x)^{\dagger}\,,\qquad {\rm
  with\,} U(\vec x) \in  
SU(2). 
\end{eqnarray}
For example, an 't~Hooft-Polyakov monopole at the origin is
described by
\begin{eqnarray}
\label{Umonopole}
U = \cos{\theta\over 2}\left(\begin{tabular}{cc}
$e^{i \psi}$ & 0 \\
0 & $e^{-i \psi}$\\ 
\end{tabular}\right) + \sin{\theta\over 2} \left( \begin{tabular}{cc} 
$0\quad$ & $i$\\
$i\quad$ & $0$\\ 
\end{tabular}\right)\,, 
\end{eqnarray}
where $0\le\theta(\vec x)\le \pi$ and $0\le\psi(\vec x)\le2\pi$ are
two parameters on the group manifold.  This choice of $U(\vec x)$
leads to the field configuration
\begin{eqnarray}
\vec\phi_1 &=&  v_1{\frac {r^i}{r}}\tau_i.
\end{eqnarray}
For this case the quantization
condition will be $Q_m g = 1.$ To get a monopole of higher charge,
we write 
\begin{eqnarray}
U_n = \cos{\theta\over 2}\left(\begin{tabular}{cc}
$e^{in \psi}$ & 0 \\
0 & $e^{-in \psi}$\\ 
\end{tabular}\right) + 
\sin{\theta\over 2} \left(\begin{tabular}{cc} 
$0\quad $ & $i$\\
$i\quad$ & $0$\\
\end{tabular}\right) \,. \\
\mbox{n = $\pm 1, \pm 2, \pm 3$, ....}\nonumber
\label{UmultiM}
\end{eqnarray}
The integer $n$ labels the homotopy class, $\pi_2(SU(2)/U(1))
\sim \pi_2(S^2) \sim Z\,,$ of the scalar field configuration.
Other choices of $U(\vec x)$ can give other configurations. For
example, a monopole-anti-monopole
configuration~\cite{Bais:1976fr} is given by the
choice
\begin{eqnarray}
\label{M-anti-M}
U = \sin{({\theta_1 - \theta_2})\over 2}\left(\begin{tabular}{cc}
0 &   $- e^{ -i \psi}$\\
$e^{i \psi}$     & 0\\ 
\end{tabular}\right) + \cos{({\theta_1 - \theta_2})\over 2}
\left( \begin{tabular}{cc}  
 $1\quad$ & $0$\\
 $0\quad$ & $1$\\ 
 \end{tabular}\right) . 
\end{eqnarray}
For our purposes, we will need to consider a $\phi_1$-vacuum
configuration with $U(\vec x) \in SU(2)$ corresponding to a
monopole-anti-monopole pair separated from each other by a distance
$> 1/v_1.$ Then the total magnetic charge vanishes, but each
(anti-)monopole can be treated as a point particle.

We also need to choose the form of the vector $\hat\kappa$ as in
Eq.~(\ref{dphi2sq}), so that it is orthogonal to $\hat \phi_1 =
\vec\phi_1/v_1$ in the internal space and rotates around $\hat
\phi_1$. Let us write
\begin{eqnarray}
\label{kappa}
\hat\kappa \equiv \hat\kappa(\vec x)^i\tau^i =
e^{i\chi(\vec x)\hat\phi_1(\vec x)}U(\vec x) 
\tau_2U^{\dagger}(\vec x)e^{-i\chi(\vec x)\hat\phi_1(\vec x)}\,.
\end{eqnarray}
We have used $\tau_2$ to write $\hat\kappa$ here but in principle
it is possible to take any constant vector orthogonal to $\tau_3$
without affecting the results below. The $\hat\phi_1(\vec x)$ used
here is constructed according to Eq.~(\ref{Uphi}) with $U(\vec x)$
as described above, and $\chi(\vec x)$ is the angle by which the
vector $U(\vec x)\tau_2U^{\dagger}(\vec x)$ is rotated in the group. 

In any case, with $\hat\kappa$ as in the above equation, we find
\begin{eqnarray}
\hat \kappa \cdot \d_\mu\hat \kappa \times \hat \phi_1 
= - \d_\mu \chi(\vec x) +  N_\mu(\vec x)\,,
\label{thetaN}
\end{eqnarray}
where the vector $N_\mu$ is given by 
\begin{eqnarray}
\label{Nvector}
N_\mu =  2i \Tr\left[ \d_\mu U U^{\dagger}\hat \phi_1\right].
\end{eqnarray}
$\chi$ is the angle which is mapped onto a circle in space to
exhibit the flux tube. As we will see now, $N_\mu$ is the (Abelian)
field of the magnetic monopoles.

Let us calculate the field strength tensor for $N_\mu$,
\begin{eqnarray}
\label{magfield}
\d_\mu N_\nu - \d_\nu N_\mu =  - \hat\phi_1\cdot\d_\mu
\hat\phi_1\times \d_\nu\hat\phi_1 +  2i\Tr[(\d_{[\mu}\d_{\nu]}U)
U^{\dagger}\hat\phi_1]\,.
\end{eqnarray}
If we use the $U(\vec x)$ of Eq.~(\ref{Umonopole}), the first term
on the right hand side of this equation is the field strength of a
magnetic monopole at the origin, while the second term is a gauge
dependent line singularity, commonly known as a Dirac string. In
this case,
\begin{eqnarray}
\label{nmu}
N_\mu = -(1+\cos\theta)\d_\mu\psi\,.
\end{eqnarray}
If $\theta$ and $\psi$ are mapped to the polar and the azimuthal
angles, $N_\mu$ is the familiar 4-potential of a magnetic monopole
with a Dirac string~\cite{Dirac:1931kp}. For the $U(\vec x)$ of the
monopole-anti-monopole pair of Eq.~(\ref{M-anti-M}), the first term
of Eq.~(\ref{magfield}) gives the Abelian magnetic field of a
monopole-anti-monopole pair, while the second term again contains a
Dirac string.

The Dirac string is a red herring, and we are going to ignore it,
for the following reason. The singular Dirac string appears because
we have used a $U(\vec x)$ which is appropriate for a point
monopole. If we look at the system from far away, the monopoles
will look like point objects, and it would seem that we should find
Dirac strings attached to each of them. However, we know that the
't~Hooft-Polyakov monopoles are actually not point objects, and
their near magnetic field is not describable by an Abelian
four-potential $N_\mu,$ so if we could do our calculations without
the far-field approximation, we would not find a Dirac string. 

There is another way of confirming that the Dirac string will not
appear in any calculations. In the far field approximation, we have
written the Lagrangian of Eq.~(\ref{vaclag}) as
Eq.~(\ref{vaclag2}), which we can rewrite using Eq.~(\ref{thetaN})
as
\begin{eqnarray}
L = - {\frac 14}\left(\d_{\mu}B_{\nu} - \d_\nu B_\mu +
  {\frac 1g}M_{\mu\nu}\right)^2 + {\frac {k^2}2} \left(gB_\mu
  -\d_\mu\chi + N_\mu\right)^2\,,
\label{stringlag}
\end{eqnarray}
where $k$ is a constant as mentioned earlier, and $M_{\mu\nu}$ is
the monopole field,
\begin{eqnarray}
M_{\mu\nu} =  - \hat{\phi}_1
\cdot\d_\mu \hat{\phi}_1\times \d_\nu\hat{\phi}_1\,.
\end{eqnarray}
The second term of the Lagrangian~(\ref{stringlag}) is the one
which exhibits a flux tube or a `physical' string (as opposed to
the unphysical Dirac string, which is an artifact of the far-field
approximation and can be relocated by a gauge transformation). An
exactly analogous term appears in the Abelian Higgs model, where
instead of $k$ we get the physical Higgs field. This non-Abelian
model also exhibits a flux string, and just like in the Abelian
Higgs model, we know that the flux string here will appear along
the zeroes of $k$, even though Eq.~(\ref{stringlag}) is written in
the far field approximation, where $k$ is a constant. The Dirac
string is also an artifact of the far-field approximation, and we
can get rid of it by choosing $U(\vec x)$ such that the Dirac
string lies along the zeroes of $k$, i.e., along the core of the
flux string. Then the troublesome line singularity, which appears
in the second term of Eq.~(\ref{stringlag}), is always multiplied
by zero, and we can ignore it for the rest of the paper.

\section{Flux tubes}
In this section we dualize the effective action obtained at
the end of the previous section, and express the theory in terms of
the string variables.  The generating functional, corresponding to
the Lagrangian of Eq.~(\ref{stringlag}), can be written as
\begin{eqnarray}
Z &=& \int \D B_\mu\D\chi\exp i
\int d^4 x \left[- {\frac 14}\left(\d_{\mu}B_{\n} 
      - \d_\nu B_\mu + {\frac 1g}M_{\mu\n}\right)^2 + 
    {\frac {k^2}2}\left(gB_\mu -\d_\mu\chi +
      N_\mu\right)^2\right]\,,
\nonumber\\
\label{flux.Higgs}
\end{eqnarray}
In the presence of flux tubes we can decompose the angle of
internal rotation $\chi$ into a regular part and a singular part,
$\chi = \chi^r + \chi^s\,.$ Then $\chi^s$ measures the homotopy
class and thus corresponds to a given magnetic flux tube, and
$\chi^r$ describes single valued fluctuations around this
configuration. The singular part of $\chi$ is related to the world
sheet $\Sigma$ of the flux string according to the equation,
\begin{eqnarray}
\epsilon^{\mu\nu\rho\lambda}\partial_{\rho}
\partial_{\lambda}\chi^s &=& \Sigma^{\mu\nu}\,, 
\label{def.sigma} \\  
\Sigma^{\mu\nu} &=&
2\pi n\int_{\Sigma}d\sigma^{\mu\nu}(x(\xi))\,\delta^4(x-x(\xi))\,,   
\label{flux.sigma}
\end{eqnarray}
where $\xi = (\xi^1, \xi^2)$ are the coordinates on the world-sheet
of the flux-tube, and $d\sigma^{\mu\nu}(x(\xi)) =
\epsilon^{ab}\partial_a x^\mu \partial_b x^\nu\,.$ In the above
equation $2\pi$ is the vorticity quantum in the units we are using
and $n$ is the winding number~\cite{Marino:2006mk}. 

Now we have integrations over both $\chi^r$ and $\chi^s\,,$ and the
second term in the action is now $\displaystyle{\frac{k^2}2}
\left(gB_\mu -\d_\mu\chi_s -\d_\mu\chi_r + N_\mu\right)^2\,.$ This
term can be linearized by introducing an auxiliary field $C_\mu\,,$
\begin{eqnarray}
&& \int \D\chi^r \exp\left[i\int d^4x {\frac {k^2}2}
\left( gB_\mu -\d_\mu\chi_s -\d_\mu\chi_r
+ N_\mu \right)^2
\right] \,
\nonumber \\ && = \int \D\chi^r \D C_\mu \exp\left[-i\int d^4 x
\left\{ \frac 1{2k^2}\,C^2_\mu + C^\mu (N_\mu + g B_{\mu} 
- \partial_\mu \chi^r - 
\partial_\mu \chi^s  )\right\}\right]\, \nonumber \\
&& = \int \D C_\mu \, \delta[\partial_\mu C^\mu]
\exp\left[-i\int d^4x \left\{ \frac 1{2k^2}\,C^2_\mu + C^\mu
(gB_\mu + N_\mu - \partial_\mu\chi^s )\right\}\right]. 
\label{flux.Cmu}
\end{eqnarray}
We can resolve the constraint $\partial_\mu C^\mu = 0$ by
introducing an antisymmetric tensor field $B_{\mu\nu}$ and writing
$C_\mu$ in the form $C^\mu = \displaystyle{\frac k2}
\epsilon^{\mu\nu\rho\lambda}\partial_\nu B_{\rho\lambda}$.

In addition, we can replace the integration over $\D\chi^s$ by an
integration over $\D x_\mu(\xi)$ which represents a sum over all
configurations of the world sheet of the flux tube. Here
$x_{\mu}(\xi)$ parametrizes the surface on which the field $\chi$
is singular. The Jacobian for this change of variables gives the
action for the string on the background space
time~\cite{Akhmedov:1995mw, Orland:1994qt}. The string has a
dynamics given by the Nambu-Goto action, plus higher order
operators~\cite{Polchinski:1991ax}, which can be obtained from the
Jacobian. We will not write the Jacobian explicitly in what
follows, but of course it is necessary to include it if we want to
study the dynamics of the flux tube. Integrating over the field
$C_\mu\,,$ we get
\begin{eqnarray}
Z = \int \D B_{\mu}&\D x_\mu(\xi)& \D B_{\mu\nu}\exp\left[i\int
d^4 x\left\{ -\frac14 F_{\mu\nu}F^{\mu\nu}  + \frac{1}{12 }
  H_{\mu\nu\rho} H^{\mu\nu\rho}  \right.\right.\,
\nonumber \\ 
&& \left.\left.+ \displaystyle{\frac k2}  \Sigma_{\mu\nu}B^{\mu\nu} 
 - {\frac k 4} 
\epsilon^{\mu\nu\rho\lambda} M_{\mu\n}     B_{\rho\lambda} 
- {\frac {k g} 2} 
\epsilon^{\mu\nu\rho\lambda} B_\mu
\partial_{\nu}B_{\rho\lambda}\right\}\right]\,,
\label{flux.Cmuout}
\end{eqnarray}
where we have written $F_{\mu\nu}= \partial_\mu B_\nu
- \partial_\nu B_\mu + \frac1g M_{\mu\nu}\,,$ defined
$H_{\mu\nu\rho} = \partial_\mu B_{\nu\rho} + \partial_\nu
B_{\rho\mu} + \partial_\rho B_{\mu\nu}\,,$ used
Eq.~(\ref{def.sigma}) and also written $M_{\mu\n} = (\d_\mu N_\n -
\d_\n N_\mu)\,.$ As explained earlier, this equality holds away
from the Dirac string, which can be ignored.

Let us now integrate over the field $B_{\mu}$. To do this we have
to linearize $F_{\mu\nu}F^{\mu\nu}$ by introducing another
auxiliary field $\chi_{\mu\nu}\,,$
\begin{eqnarray}
&& \int \D B_{\mu} \exp \left[ i\int d^4x \left\{- \frac14
F_{\mu\nu}F^{\mu\nu}-   
\frac {k g}2 \epsilon^{\mu\nu\rho\lambda}B_\mu
\partial_{\nu}B_{\rho\lambda} - {\frac k 4} 
\epsilon^{\mu\nu\rho\lambda} M_{\mu\n} B_{\rho\lambda} \right\} 
\right] \,
  \nonumber\\ 
&& = \int \D B_{\mu} \D \chi_{\mu\nu} \exp\left[i\int d^4x 
\left\{- \frac14 \chi_{\mu\nu}\chi^{\mu\nu} + \half
\epsilon^{\mu\nu\rho\lambda} \chi_{\mu\nu}
\partial_\rho B_\lambda  + {\frac 1{4g}}
\epsilon^{\mu\nu\rho\lambda} \chi_{\mu\nu}M_{\rho\lambda}  
\right.\right.\,
\nonumber \\ 
&& \phantom{xxxxxxxxxxxxxxxxxxxxxxxxx} \left.\left. 
- {\frac {k g}2}  
\epsilon^{\mu\nu\rho\lambda}
B_{\mu\nu}\partial_\rho B_\lambda - {\frac k 4} 
\epsilon^{\mu\nu\rho\lambda} M_{\mu\n}     B_{\rho\lambda}
\right\}\right]\,  
\nonumber \\  
&& = \int \D
\chi_{\mu\nu}\,\delta\Big[\epsilon^{\mu\nu\rho\lambda} 
\partial_{\nu}(\chi_{\rho\lambda}-gk
B_{\rho\lambda})\Big]\,\nonumber\\  
&&\phantom{xxxxxxxxxxxxxx}\exp\left[i\int d^4x\left\{ -\frac 14 
\chi_{\mu\nu}\chi^{\mu\nu} + 
 {\frac 1{4g}}  \epsilon^{\mu\nu\rho\lambda}   (\chi_{\mu\nu} - 
k g B_{\mu\n})M_{\rho\lambda}\right\}\right]\,.
\label{flux.Alinear}
\end{eqnarray}
We can integrate over $\chi_{\mu\nu}$ by solving the
$\delta$-functional as 
\begin{eqnarray}
\chi_{\mu\nu} = gk B_{\mu\nu} + \partial_\mu A_\nu^m - \partial_\nu
A_\mu^m\,. 
\label{flux.Adual}
\end{eqnarray}
Here $A^m_{\mu}$ is the `magnetic photon', and what we have
achieved is dualization of the vector potential $B_{\mu}$. The
result of the integration is then inserted into
Eq.~(\ref{flux.Cmuout}) to give
\begin{eqnarray}
Z =&& \int \D A^m_{\mu} \D x_{\mu}(\xi) \D B_{\mu\nu}\nonumber\\
&&\exp\left[i\int\left\{- 
\frac 14 \left(gk B_{\mu\nu} + \partial_{[\mu}A^m_{\nu]}\right)^2
+ \frac 1{12 } H_{\mu\nu\rho}H^{\mu\nu\rho} - 
{\frac k 2} \Sigma_{\mu\nu}B^{\mu\nu} - j_m^{\m}A^{m\mu}
\right\}\right].  
\label{flux.functional}
\end{eqnarray}
Here $j_m^{\mu} = - {\frac 1{2g}} \epsilon^{\mu\nu\rho\lambda}\d_\n
M_{\rho\lambda}$ is the current of magnetic monopoles. The
equations of motion for the field $B_{\mu\nu}$ and $A^{\mu}$ can be
calculated from this to be
\begin{eqnarray}
\label{flux.Beom}
\partial_\lambda H^{\lambda\mu\nu} &=& -m \, G^{\mu\nu} -
\frac{m}{g} \,\Sigma^{\mu\nu} \,,\\
\d_\mu G^{\mu\n} &=& j_m^\mu
\label{flux.Aeom} 
\end{eqnarray}
where $G_{\mu\nu}= gk B_{\mu\nu} + \partial_{\mu}A^m_{\nu} -
\partial_{\nu}A^m_{\mu}\,,$ and $m = gk$. Combining
Eq.~(\ref{flux.Aeom}) and Eq.~(\ref{flux.Beom}) we find that
\begin{equation}
\frac 1g  \partial_\mu \Sigma^{\mu\nu}(x) + j_m^\mu(x) = 0\,.
\label{mono.coneq}
\end{equation}
It follows rather obviously that a vanishing magnetic monopole
current implies $\partial_\mu \Sigma^{\mu\nu}(x) = 0\,,$ or in
other words if there is no monopole in the system, the flux tubes
will be closed. 

If there are monopoles at the ends of the flux tube, we need to
check if the fluxes match correctly. The magnetic flux through the
tube is $\frac{2n\pi }g\,,$ while the total magnetic flux of the
monopole is $\frac{4m\pi}g\,,$ where $n, m$ are integers. So in
order for the monopoles to be confined, we must have $n = 2m\,.$ It
follows that the minimum value of $n$ required for confinement of
't Hooft-Polyakov monopoles is $n=2\,.$ Clearly, if we could create
a monopole-anti-monopole pair, it could break the flux tube. But we
have a hierarchy of energy scales $v_1\gg v_2\,,$ which are
respectively proportional to the mass of the monopole and the
energy scale of the string. We can expect this hierarchy to prevent
pair creation from string breakage.

The conservation law of Eq.~(\ref{mono.coneq})
also follows directly from the gauge invariance of the action in
Eq.~(\ref{flux.functional}) under the gauge transformations
\begin{eqnarray}
\delta B_{\mu\nu} = \partial_\mu\Lambda_\nu
- \partial_\nu\Lambda_\mu\,, \qquad \delta A^m_\mu = -
gk\Lambda_\mu\,. 
\end{eqnarray}
Applying this gauge transformation to the action and integrating
over all $\Lambda_\mu\,,$ we can write the partition function
as~\cite{Chatterjee:2006iq} 
\begin{eqnarray}
Z =&&\int \D A^m_{\mu} \D x_{\mu}(\xi) \D B_{\mu\nu}
\delta\Big[\frac 1g  
\partial_\mu\Sigma^{\mu\nu} +  j^\nu_m\Big] \nonumber\\
&&\exp\left[i\int\left\{- 
\frac 14 (gk B_{\mu\nu} + \partial_{[\mu}A^m_{\nu]})^2
+ \frac 1{12} H_{\mu\nu\rho}H^{\mu\nu\rho} - 
{\frac k 2} \Sigma_{\mu\nu}B^{\mu\nu} - j_m^{\mu}A^{m\mu}
\right\}\right] \,.  
\end{eqnarray}
Now we can define a gauge invariant $B'_{\mu\n} = B_{\mu\nu} +
{\frac 1{gk} } (\partial_\mu A^m_\n - \partial_\n A^m_\mu)$.  The
generating functional becomes,
\begin{eqnarray}
\label{confinement}
Z =\int \D x_{\mu}(\xi) \D B'_{\mu\nu}&&
\delta\Big[\frac1g\partial_\mu\Sigma^{\mu\nu}(x)  
+  j^\nu_m(x)\Big] \nonumber\\
&&\exp\left[i\int\left\{\frac 1{12} H_{\mu\nu\rho}H^{\mu\nu\rho} 
- \frac 14  m^2 {B'}^2_{\mu\nu}- 
{\frac m{2g}} \Sigma_{\mu\nu}{B'}^{\mu\nu} \right\}\right] \,. 
\end{eqnarray}
Thus these strings are analogous to the confining strings in three
dimensions~\cite{Polyakov:1996nc}. There is no $A^m_\mu$, the only
gauge field which is present is $B'_{\mu\n}$. This $B'_{\mu\n}$
field mediates the direct interaction between the confining
strings.

The functional delta function at eq.~(\ref{confinement}) enforces
that at every point of space-time, the monopole current cancels the
currents of the end points of flux tube.  So the monopole current
must be non-zero only at the end of the flux tube.
Eq.~(\ref{confinement}) does not carry Abelian gauge field
$A^m_{\mu}$, only a massive second rank tensor gauge field.  All
this confirms the permanent attachment of monopoles at the end of
the flux tube which does not allow gauge flux to escape out of the
flux tubes.


\end{document}